# Competitive and Cooperative electronic states in $Ba(Fe_{1-x}T_x)_2As_2$ with T=Co, Ni, Cr

Qiang Zou[1], Mingming Fu[1,2], Zhiming Wu[2], Li Li[3], David S. Parker[3], Athena S. Sefat[3], and Zheng Gai[1,*]

[1]Center for Nanophase Materials Sciences, Oak Ridge National Laboratory, Oak Ridge, TN 37831, USA

[2]Fujian Provincial Key Laboratory of Semiconductors and Applications, Collaborative Innovation Center for Optoelectronic Semiconductors and Efficient Devices, Department of Physics, Xiamen University, Xiamen, Fujian Province, 361005, P. R. China

[3]Materials Science & Technology Division, Oak Ridge National Laboratory, Oak Ridge, TN 37831, USA

[*] Corresponding author: gaiz@ornl.gov

## Abstract

The electronic structure inhomogeneities in Co, Ni, and Cr doped $BaFe_2As_2$ '122' single crystals are compared using scanning tunneling microscopy/spectroscopy (STM/S) at the nanoscale within three bulk property regions in the phase diagram: a pure superconducting (SC) dome region (Co-122), a coexisting SC and antiferromagnetic (AFM) region (Ni-122), and a non-SC region (Cr-122). Machine learning is utilized to categorize the various nanometer scale inhomogeneous electronic states, described here as in-gap, L-shape and S-shape states immersed into the SC matrix for Ni-and Co-doped 122, and L-shape and S-shape states into the metallic matrix for Cr-doped 122. Although the relative percentages of in-gap, L-shape and S-shape states vary in the three samples, the total volume fraction of the three electronic states is quite similar. This is coincident with the number of electrons ($Ni_{0.04}$ and $Co_{0.08}$) and holes ($Cr_{0.04}$) doped into the 122 compound. By combining the volume fractions of the three states, the local density of states (LDOS), magnetic field dependent behavior and global properties in these three samples, the in-gap state is confirmed as a magnetic impurity state from the Co or Ni dopants. In addition, the L-shape state is identified as a spin density wave (SDW) which competes with the SC phase, and the S-shape state is found to be another form of magnetic order which constructively cooperates with the SC phase, rather than competing with it. The comparison of the vortex structures indicates that the inhomogeneous electronic states serve as pinning centers for stabilizing the vortex lattice.

**Keywords:** Iron-based superconductor, magnetic dopant, scanning tunneling microscopy, spectroscopy, electronic inhomogeneity, spin density wave, arsenides, 122

## Introduction

The interplay between magnetism and superconductivity (SC) is one of the fundamental topics for understanding the mechanism of superconductivity of unconventional superconductors in the cuprates and iron-based superconductors (FeSCs) [1,2], since SC appears near antiferromagnetic (AF) order. In cuprates, experiments demonstrated that the superconducting pairing state is the unconventional $d_{x2-y2}$, which is believed induced by spin fluctuations [3-5]. Similarly to the cuprates, magnetism in the FeSCs plays an important role in the electron pairing mechanism [2,4]. However, in the FeSC it is more complicated and controversial since Fe 3d orbitals form multiple Fermi surfaces (FS), while in cuprates only a single Cu d-band crosses the FS [6-9]. For example, in the most studied FeSC family of pnictides $BaFe_2As_2$ (122), SC can be realized by atomic substitution at any element site, for example by hole-doping onto Ba site[10], electron-doping on Fe site [11] [12] and isovalent doping on As site[13]. However, it is intriguing that crystals with electron-doping to Fe site (with Ni and Co) are superconducting while crystals with hole-dopants (with Cr) are not [14].

On the other hand, for the electron-doped 122 crystals, it is well established that upon electron doping via Co/Ni substitution for Fe, the collinear long-range AF order is suppressed, and SC then appears. However, there are ongoing debates concerning the relationship between the AF order and SC. One perspective arises from a presumed itinerant nature of magnetism. In this view the static AF order arises from the formation of a spin-density-wave induced by itinerant electrons and Fermi surface nesting of the electron and hole pockets. Upon doping the pair scattering between the electron and hole like FS pockets then leads to SC[15,16]. A different perspective arises from a presumed localized nature of magnetism: the short-range incommensurate AF order is a cluster spin glass phase arising from disordered localized moments[12,17]. There is an additional perspective that local moments and itinerant electrons *coexist*. In this scenario part of the Fe d bands are delocalized and contribute to the itinerancy, whereas others are localized due to a strong correlation effect and provide the source for local moments[18,19]. This model was used to explain the scanning tunneling microscopic observation of the coexistence as well as the co-disappearance of a pseudogap-like feature and superconductor in the $NaFe_{1-x}Co_xAs$ (Co-111) system where the pure itinerant picture apparently fails [20]. A muon spin rotation study also observed a spatially inhomogeneous magnetic state developing below $T_c$ which, surprisingly, has a constructive relationship with SC in near optimal or overdoped Co-122 samples [21,22]. This is intrinsically different from the reported SDW phase in $NaFe_{1-x}Co_xAs$ (Co-111) which competes with, and is
anti-correlated, with the SC phase [23].

Here, we investigate the local electronic structure of 122 crystalline matrix $BaFe_2As_2$ that
is doped with magnetic elements (Ni, Co, or Cr) at the same amount of electron versus hole doping
levels per Fe atom. We systematically explore the electronic structure signatures of the magnetism
and SC phases using low temperature high magnetic-field scanning tunneling
microscopy/spectroscopy (STM/S). The machine learning technique of K-means clustering
method is employed to categorize the various nanometer-size inhomogeneous electronic states. In
addition to the SC state, an in-gap state, a competitive L-shape and a cooperative S-shape state are
found to coexist in the samples. The in-gap state in the SC crystals is confirmed to be a magnetic
impurity state from Co or Ni dopants, while the L-shape state is identified as an SDW that
competes with the SC phase. In addition, the S-shape state originates from a local magnetic order
that surprisingly, constructively cooperates with the SC phase rather than competing with it. A
comparison of the vortex structures indicates that those inhomogeneous electronic states serve as
pinning centers for stabilizing the hexagonal vortex lattice for the bulk superconductors.

## Results

For the purpose of comparison, we have studied three sets of samples of the transition-
metal substituted $Ba(Fe_{1-x}T_x)_2As_2$ in three different bulk property regions (bulk SC, magnetism
and SC coexistence, and non-SC). These are the bulk SC $Ba(Fe_{0.92}Co_{0.08})_2As_2$ (Co-122) crystal,
the AFM and SC coexisting $Ba(Fe_{0.96}Ni_{0.04})_2As_2$ (Ni-122) crystal, and the AFM
$Ba(Fe_{0.96}Cr_{0.04})_2As_2$ (Cr-122) crystal. Since Ni (Cr) doping introduces twice the number of
electrons (holes) in the FeAs layer as that of Co doping (from simple electron counting), we select the Co-122 doping level as 0.08 instead of 0.04 to achieve the same amount of electron versus hole
doping levels per Fe atom. This ensures consistency across the three sets of samples. Cr-122 has a
G-AFM ground state ($T_N$~ 100 K), while optimally doped Co-122 is in the SC state ($T_c$ ~ 22 K)
[24]; and the composition of the Ni-122 was chosen to locate it in the coexistence range of SC and
AFM orders ($T_c$ ~ 19 K, $T_N$~ 45 K). All the scanning tunneling microscopy/spectroscopy
experiments are performed at the base temperature of ~ 4 K. These phases are indicated using blue,
red and yellow balls in the temperature-composition (T-x) phase diagram in Fig. 1(a); the crystal
structure of the undoped parent 122-unit cell is in inset.

After low-temperature cleavage, the three sets of samples show similar topography, which
is the coexistence of 2×1 and √2 × √2 reconstructions, as reported for Co-122 before [25,26]. Fig. 1(b)
shows a typical large-scale topographic image of Ni-122, with terraces, steps in 0.7 nm (half unit
cell in *c* axis) and defects. The atomic resolution image collected from the red box of Fig. 1(b)
reveals the √2 × √2 reconstruction, atomic scale dark pits, protrusions and dark lines, as shown in
Fig. 1(c). The dark lines are the antiphase boundaries of the √2 × √2 domains, the dark pits are the
results of missing single or multiple Ba atoms on the top surface, and the bright protrusions reflect
the electronic contributions of the elemental defects beneath the Ba surface based on the height of
the protrusion (~ 0.04 nm). Note that there exist height-variations of approximately 20 pm even
on atomically well-ordered areas (~5 pm atomic corrugation from relative homogeneous areas),
suggesting the existence of electronic inhomogeneity in the sample.

Dramatic local electronic inhomogeneity is found on the Ni-122 surface. Fig. 1(d) presents
the STS spectra along the black arrow in Fig. 1(c). Although the morphology of the areas around
the arrow is similar (√2 × √2, no defects), the STS spectra, which are proportional to the local
density of states (LDOS), show remarkably different features. The spectra along the black arrow exhibit not only superconducting coherent peaks with an SC gap width of about 4.5 meV in the
blue curves, but also an in-gap state around the Fermi level, with the suppression of the SC
coherence peaks in the brown curves.

To clarify the correlation between the topography and the local electronic inhomogeneity
on Ni-122, current imaging tunneling spectroscopy (CITS) maps were collected from various areas
on the surface. Systematic comparison among topography, CITS, and SC gap map concludes that
there is no apparent correlation between the topography and local electronic inhomogeneity. Such
a comparison from an area with fairly ordered $\sqrt{2} \times \sqrt{2}$ reconstruction and few defects (blue box
in Fig. 2(a)) is shown in Fig. 2. Shown in Fig. 2(b) and 2 (c) are the STS map at Fermi level and
the corresponding gap map. The SC gap map in this work is the map of the SC gap value at each
pixel, where the value of Δ is obtained by fitting each dI/dV using Dynes model of $\frac{dI(V)}{dV} =$
$\int_{-\infty}^{\infty} dE \, \mathrm{Re}\left\{\frac{E-i\Gamma}{\sqrt{(E-i\Gamma)^2-\Delta^2}}\right\}\left\{\frac{1}{k_BT}\frac{(E+eV/k_BT)}{[1+\exp(E+eV/k_BT)]^2}\right\}$, as we did previously [25]. Patches of non-SC areas
(green) are immersed in the SC matrix (blue to white area, with various SC gap widths). The
nanometer size inhomogeneity in the STS map and gap map are highly correlated to each other,
but not with the almost homogeneous topographic image. To reveal the intrinsic electronic
structure of the inhomogeneity, an extensive survey of the spectra on the surface is conducted.
Other than the SC gap features, two other types of characteristic STS are found on the Ni-122
surface, also noncorrelated to the topography. Two such areas are circled in blue and brown in Fig.
2(b). Line STS in Fig. 2(d) and (e) along the dashed green and solid brown arrows in Fig. 2(b)
reveal the electronic structures surrounding the inhomogeneity areas, changing gradually from SC
(blue curves) to non-SC (brown or green curves). The brown curves in Fig. 2(d) show the
suppression of the SC coherence peak by an in-gap state around Fermi level. The suppression area extends about 1 nm from the center (brown circle in Fig. 2(b)), consistent with the observation of
the magnetic impurity induced in-gap state in Co-122 [25]. The green curves in Fig. 2(e) have an
elevated density of states at Fermi level, which is named as an S-shape state in this work.

In order to analyze the spatial distribution of all three types of spectra, K-means clustering
analyses of the spectra of CITS in Ni-122 are performed. K-means clustering is a vector
quantization data mining approach, which groups a large dataset into components with principal
characteristics [27]. Fig. 2(f) shows the three principal responses in this data set, which correspond
to an SC, an in-gap and an S-shape state via comparison with the STS in Fig. 2(d) and (e). In the
cluster map, each color represents the spatial distribution of the corresponding principal dI/dV
spectra. As we can see, the cluster map captures the transition between the phases in dI/dV line
mappings across the boundaries (Fig. 2(d), and (e)). Furthermore, the distribution of the SC state
in the K-means clusters is the same as the distribution of the STS map and gap map in Fig. 2(b)
and (c). Most importantly, the non-SC areas are clearly categorized into distinct areas of an in-gap
state (brown) and an S-shape state (green).

The above coexistence of S-shape and in-gap states within superconducting crystals are
not affected by the surface termination since they are observed not only on a $\sqrt{2} \times \sqrt{2}$ reconstruction
but also on $2 \times 1$ reconstructed areas. It has been reported that the $\sqrt{2} \times \sqrt{2}$ and $2 \times 1$ surface
reconstructions coexist on the cleavage surface of $BaFe_2As_2$[25,26]. However, the superconductivity
gap width is not affected by the surface reconstructions owing to the global nature of the
superconductivity [25]. These observations strongly suggest that the S-shape, in-gap state and
superconducting state are in fact *intrinsic* electronic structures of the superconducting 122 system
rather than the surface specific properties.

The successful grouping of the characteristic states from large datasets of CITS in Ni-122
suggests the usage of the K-means clustering analysis a convenient tool for a statistical comparison
of the transition metal substituted $Ba(Fe_{1-x}T_x)_2As_2$, i.e., Ni-122, Co-122 and Cr-122. Shown in Fig.
3(a), (d) and (g) are typical large-size STM topographic images of terraces in well cleaved samples.
Using the above K-means clustering method, the spatial distribution, as well as the corresponding
principal responses of Ni-122, Co-122, and Cr-122, are presented in Fig. 3(b) and (c), Fig. 3(e)
and (f) and Fig. 3(h) and (i), respectively. Also, the corresponding principal responses with shaded
error bars (spectral distribution plots) are shown in Fig. S1. Note that, in addition to SC, in-gap
and S-shape states, and L-shape states are observed in all crystals. The L-shape states are presented
as yellow curves in Fig. 3(c), (f) and (i), a group of line dI/dV spectra in Fig. S2 shows the
crossover between SC and L-shape state. The dominant domains in Fig. 3(b), (e) and (h) are the
percolated blue areas, which corresponds to the blue curves in Fig. 3(c), (f) and (i) as SC state in
Ni-122 and Co-122, and to a metallic state in Cr-122, respectively, consistent with the bulk
physical properties of the samples.

All those samples share a substantial volume of green and yellow areas in the blue matrix,
which correspond to the S-shape state (green curves) and L-shape state (yellow curves) in Fig. 3(c),
(f) and (i). In Fig. 3(b) and (e), both superconducting Ni-122 and Co-122 surfaces have brown
areas which correspond to the in-gap state (brown curves) in Fig. 3(c) and (f), while in Fig. 3(g)
the in-gap state is not observed on the non-superconducting Cr-122 surface. Although the energy
scale of the spectral feature for the S-shape and L-shape states for Ni, Co-122 vs Cr-122 varies
slightly, the characteristic features are the same. For the S-shape there is a highly elevated DOS at
Fermi level, and a local minimum at filled state, while for the L-shape there is a much lower DOS
at the Fermi level, and a local minimum at empty state. We speculate that the differences in the energy scales might be the result of the influence of the superconducting matrix in Ni-122 and Co-
122 which is notably absent in Cr-122. Although the S-shape, L-shape and in-gap states are
immersed in the SC (or metallic for Cr-122) matrix, the spatial distribution of those states is not
correlated. Fig. 3(j) shows the area percentages of each principal response in Ni-122, Co-122 and
Cr-122 from the three sets cluster maps in Fig.3. The total volume of the majority domains (SC or
metallic) is similar for the three sets, but the volumes of the S-shape and L-shape states show a
noticeable difference between the SC (Ni-122 and Co-122) and metallic (Cr-122) compounds. In
Cr-122, the percentage of L-shape state is 2.4 times higher than in Ni-122(Co-122), while the S-
shape state is rarely found. Although the relative percentage of S- or L- shape states are different,
the total volume of the two states is quite similar in the three compounds, despite the notable
differences that Cr-122 is non-SC, Ni-122 is an underdoped SC and Co-122 is an optimally doped
SC, coincident with the same amount of hole ($Cr_{0.04}$) / electrons ($Ni_{0.04}$ and $Co_{0.08}$) doped into the
three compounds. It is important to note that while the distribution of those states in Co-122 and
Cr-122 is quite uniform, Ni-122 shows strong spatial variations. In Ni-122, there are phase
segregated areas with an almost pure SC state, and areas with four highly mixed electronic states.
As shown from the spectral distribution plots (Fig. S1), the overlap among different states is small.
Furthermore, fairly large size CITS images are used for the analysis instead of small size ones,
leading to better capturing not only the influence of the local inhomogeneity but also the average
features of the surface.

Fig. 3(k) shows a direct comparison of the average LDOS of the in-gap state in Ni-122 and
Co-122 (the corresponding spectral distribution is shown in Fig. S3). The in-gap state of Ni-122
samples shows a higher density of states around the Fermi level and superconducting coherence
peaks with an asymmetrical suppression. In view of the different form of the in-gap states in two compounds (Ni-122 and Co-122) and the observation of the zero-energy peak state only in the Ni-122 and Co-122 crystals (never in Cr-122 ), we conclude that the robust in-gap state in the iron-based 122 samples originates from the Co and Ni dopants, rather than from Fe defects [25]. Included in the Supplemental Information (Fig. S4) are density functional theory calculations of the effects of these dopants on the $BaFe_2As_2$ DOS, which find a substantial dopant-associated DOS variation even at rather dilute concentration. While these non-magnetic calculations cannot be directly compared with experimental data from the superconducting state, they are nonetheless suggestive of the ability of even small dopant concentrations to substantially alter materials properties – as the very existence of superconductivity in this material class reflects.

To reveal the origin of various states, the samples were investigated under a magnetic field applied perpendicular to the cleavage surfaces. Under magnetic field, the superconducting state transforms into the Abrikosov state by forming a vortex structure, where non-superconducting cores are surrounded by superconducting areas. Such a vortex structure can be imaged by spatial STS mapping around the Fermi level, with higher local density of states bright spots in the cores [28-30]. Shown in Fig. 4 (a), (b) and (c) are the STS maps at Fermi level (zero bias) under a perpendicular magnetic field of 2, 4 and 6 T, respectively. These maps were collected from the same location on the Ni-122 surface. STS curves across a vortex core are plotted in Fig. 4(m) to demonstrate the spatial evolution of the tunneling conductance. Approaching the core center, the zero bias conductance increases, and the superconducting coherence peaks are suppressed. Consistent with previous observations from electron-doped 122, no Andreev bound state is observed up to 6 T in the vortices of Ni-122 [30,31]. The Ginzburg-Landau coherence lengths and the vortex lattice constants extracted from the normalized, azimuthally averaged radii of average vortices in the STS maps and their self-correlation images are 4.8, 4.5, 4.1 nm and 33.7, 24.0, 19.5 nm at 2, 4 and 6 Tesla, respectively. The density of vortices increases linearly with the magnetic
field as shown in Fig. 4(n). By fitting the applied magnetic field dependent vortex densities, the
vortex magnetic flux is extracted as $\Phi = (1.96 \pm 0.4) \times 10^{-15}$ $Tm^2$, which is in excellent agreement
with the single flux quantum of $\Phi_0 = 2.07 \times 10^{-15}$ $Tm^2$ [28-30].

The symmetry of vortex lattices changes dramatically with the increase of the magnetic
field. Figs. 4(d), (e) and (f) show two-dimensional autocorrelation images calculated from the
vortex images in (a), (b) and (c). The vortex forms a weak triangular lattice at 2 T with a six-fold
correlation shown in Fig. 4(d). At higher fields, the vortex lattice gradually changes to a rectangular
shape at 4 T (Fig. 4(e)) and becomes a more isotropic ring-like structure at 6 T (Fig. 4(f)). The six-
folded symmetry and ring-like pattern can be faintly seen in the two dimensional fast Fourier
transformation (FFT) images in Figs. 4 (g) and (i), and the rectangular symmetry in the 4T image
is not recognizable in Fig. 4 (h). The 2D autocorrelation image presents more information in the
vortex lattice than the FFT image since the former extract the spatial correlation among all vortices
and so is sensitive to the anisotropy of the features, while the latter decomposes the signal into its
harmonic components which is usually applied to ordered lattices. The emergence of a 4-fold
symmetry on top of 6-fold symmetry suggests that the configuration of the vortices is affected by
the underlying 4-fold crystal lattice. The emergence of the ring-like structure also suggests that the
distance correlation among the vortices is stronger than the orientational correlation in Ni-122.
The Delaunay triangulation analysis is shown in Fig. 4(j), (k) and (l). The defect vortices are
symbolized by blue and red triangles, which have 5 and 7 nearest neighbors rather than 6 in regular
vortex lattice.

It is noteworthy to point out that although the vortex densities for the two compounds are
almost the same, the vortex matter for Ni-122 and Co-122 is different. In contrast to the gradual change of the vortex lattice in Ni-122, the vortex lattice shows no change in Co-122 under a
magnetic field up to 6 T. This difference is also reflected in the different trends of defect-vortex
fractions of the two compounds, as shown in Fig. 4(o) and Fig. S5 in the Supplementary
Information. While the vortex matter of Ni-122 is quite similar to LiFeAs [29], its behavior in Co-
122 is more similar to (BaK)$Fe_2As_2$ [31]. We anticipate that the different vortex matters for Ni-122
and Co-122 relate to the drastically different phase distributions in the two compounds: uniformly
scattered S, L and in-gap states in the Co-122 compound serve as pinning nuclei for stabilizing the
hexagonal vortex lattice, consistent with the observation of surface pinning effect in (BaK)$Fe_2As_2$
sample[31]. This observation suggests that the inhomogeneous electronic states play an important
role in the higher upper critical fields in the FeSC.

The magnetic field influence on all states is summarized in Fig. 5, which shows the result
from Co-122 under magnetic field up to 6 Tesla. In this plot, each set of field dependent spectra is
extracted from the same location. The gradually increased zero bias conductance and suppressed
superconducting coherence peaks under magnetic field shown in Fig. 5(a) were discussed in the
previous few paragraphs. For the in-gap state, the peak is pinned at the Fermi level even up to 6
Tesla, but the intensity of the peak increases slightly with the magnetic field as shown in Fig. 5(b).
A similar robust zero-energy bound state was also observed on Fe(Te,Se)[32] which in that material
is caused by interstitial Fe impurities.

For both the S-shape and L-shape states, there is no significant change detected up to 6 T
as shown in Fig. 5(c) and (d). These observations exclude observation of the Kondo effect of
magnetic moment in metal/superconductor since no splitting was found in the magnetic field
dependent spectra. Although the S-shape and L-shape spectra look similar to the Fe-vacancy-
induced bound state found in the K-doped iron selenide [33], their extremely weak field dependence indicates that either the g-factor in our case is far smaller than 2.1 or that these spectra are from a
different origin.

The L-shape like state has been identified as a spin density wave spectrum with the
signature features of a large asymmetry with respect to the Fermi level and a large residual DOS
at $E_F$ [20,23]. This is found both coexisting and competing with SC states in under- and over- doped
$NaFe_{1-x}Co_xAs$ (Co-111) FeSC. In our current comparison of the 122 system, we further found that
1) the SDW spectrum exists in both electron and hole doped 122 compounds; 2) the SDW survives
even in the optimally doped Co-122 crystals, although the volume is very small (2.3%); 3) the
highly asymmetric spectrum results from the usual particle-hole symmetry being tilted towards
positive bias, similarly to the positive bias in Co-111 case. These results suggest that the
coexistence of local moments and itinerant electrons, as modelled for the Co-111 system [20,23]
should also hold for the 122 system: Cooper pairing can occur when portions of the Fermi surface
are already gapped by the SDW order. The much higher percentage of the L shape state in Cr-122
offers a hint as to why there is no SC in the Cr-122. Further variable temperature experiments are
necessary here to confirm the SDW gap and the transition temperature.

The S shape state was not reported previously. Although the present data set cannot firmly
prove the origin of this phase, we believe it is another form of magnetic state, based on its spectral
features. The main differences between the S and L shape states are: 1) the asymmetric spectrum
is tilted to the negative bias; 2) the Fermi surface residual DOS of the L shape is low, very close
to the bottom of SC gap, but for S shape the Fermi surface residual is much higher (Fig. 3 (c) (f)
and (i)); 3) the volume percentage of the S state is lowest in non-SC Cr-122, and highest in the
optimally doped Co-122. Notably, from the volumes of the S state of the three compounds, the
behavior of the S shape state is more cooperative with the superconducting state, rather than competitive with SC as observed for the L shape state. This abnormal behavior and the highly elevated LDOS at Fermi level suggest the S-shape state should have the same origin as in the muon spin rotation-based observation of a different component of spin density wave order in Co-122 [21,22] and the neutron study of the identification of a short-range spin-glass state in Ni-122 [12]. In the muon spin rotation study, a spatially inhomogeneous magnetic state develops below $T_c$ and it has a constructive rather than a competitive relationship with SC near optimally or even overdoped Co-122 samples [21,22]. A follow-up theoretical study explains the distinct magnetic phases using an impurity-induced order which is stabilized by a multiple-Co dopant effect [34]. Future spin-polarized STM studies are essential to further determine the relations between the S and L shape states as well as that of the in-gap state with the superconductivity phase.

## Conclusion

In summary, the nanoscale electronic structures of $Ba(Fe_{1-x}T_x)_2As_2$ single crystals in three different phases of optimally-doped SC Co-122, coexisting SC and AFM Ni-122, and non-SC Cr-122, are studied with STM/S at 4 K. All these crystals show nanometer-size local electronic inhomogeneities that are unrelated to the surface topography. The total volume of the inhomogeneous areas is similar in the three compounds, coincident with the same amount of number of electrons ($Ni_{0.04}$ and $Co_{0.08}$) / holes ($Cr_{0.04}$) doped into the three crystals. But the distribution of the inhomogeneous electronic states is different in that they are uniformly distributed all around the sample in nanometer-size clusters in optimally-doped Co-122 crystals, but more segregated in under-doped Ni-122. Using a K-means clustering statistical method, the local electronic states are categorized as S-shape, L-shape and in-gap states immersed in the matrix of SC state for Ni-122 and Co-122, and S-shape and L-shape states in the metallic phase for Cr-122. The spatial distribution of S-shape, L-shape and in-gap states are not correlated. The comparison of vortex structures under different magnetic fields in areas with pure SC state or mixed inhomogeneous areas indicates that the S-shape, L-shape and in-gap states serve as pinning nuclei for stabilizing the hexagonal vortex lattice. By combining all the observations including the volume fractions of the three states, LDOS, field dependent behavior and global properties in the three sets of samples, we confirm the following. Firstly, the in-gap state in SC crystals originates from the magnetic Co/Ni dopants. Secondly, the origin of the L-shape state is an SDW that competes with the SC phase. Thirdly and most compellingly, another form of magnetic order observed in an S-shape state constructively cooperates with the SC phase or serves as an essential element for SC.  This research also demonstrates the usage of machine learning to categorize spectra as a powerful approach to analyze hyperspectral dataset, and aid in uncovering out meaningful physics beneath it.

## Method

Single crystals of $Ba(Fe_{0.92}Co_{0.08})_2As_2$, $Ba(Fe_{0.96}Ni_{0.04})_2As_2$ and $Ba(Fe_{0.96}Cr_{0.04})_2As_2$ were grown out of self-flux from FeAs and Co (or Ni, Cr)As binaries, a synthesis method similar to our previous reports [24]. Phase purity, crystallinity, and the atomic occupancy of all crystals were checked by collecting Powder X-ray diffraction (XRD) data on an X'Pert PRO MPD diffractometer (Cu $K_{\alpha 1}$ radiation, $\lambda$=1.540598 Å); The average chemical composition of each crystal was measured with a Hitachi S3400 scanning electron microscope operating at 20 kV, and use of energy-dispersive x-ray spectroscopy (EDS).

All crystals were cleaved in ultra-high vacuum (UHV) at ~ 78 K and then immediately transferred to the Scanning Tunneling Microscopy/Spectroscopy (STM/S) head which was precooled to 4.2 K without breaking the vacuum. The STM/S experiments were carried out at 2.0
K or 4.2 K using a UHV low-temperature and high field scanning tunneling microscope with base
pressure better than $2\times10^{-10}$ Torr [25]. Pt-Ir tips were mechanically cut then conditioned on clean Au
(111) and checked using the topography, surface state and work function of Au (111) before each
measurement. The STM/S were controlled by a SPECS Nanonis control system. Topographic
images were acquired in constant current mode with bias voltage applied to sample, and tip
grounded. All the spectra were obtained using the lock-in technique with a modulation of 0.1 to 1
mV at 973 Hz on bias voltage. Line spectroscopies and Current-Imaging-Tunneling-Spectroscopy
were collected over a grid of pixels at bias ranges around Fermi level using the same lock-in
amplifier parameters. During STM/S measurement, a magnetic field up to 9 Tesla was applied
perpendicular to the crystal surfaces.

The K-means clustering analysis uses an unsupervised learning algorithm, with the goal to
find groups in the data, with the number of groups represented by the variable K. K-means method
groups a large dataset into components with quintessential characteristics. K-means method itself
is a well-developed method in the community. We use the K-means codes from PYCROSCOPY
package [27]. Our own software package was developed by utilizing the K-means clustering method
to minimize cognitive bias during processing large amounts of dI/dV mapping data (CITS image).
More details in supplementary information.

Calculations were performed using the commercially available plane-wave density
functional theory-code WIEN2K [35] employing the generalized gradient approximation of Perdew,
Burke and Erzenhof [36]. Convergence of the calculations was carefully checked, both with respect
to RKmax (the product of the smallest sphere and largest plane-wave expansion wavevector, here
taken as 8.0) and the number of k-points, with 1000 being used in the full Brillouin zone). Spin-orbit coupling was not included, and no internal coordinate relaxation was performed. All calculations used the low-temperature orthorhombic experimental structure of $BaFe_2As_2$ by Avci et al [37] in a 2 x 2 x 1 supercell of this structure, comprising 80 atoms total, including 32 transition metal atoms. Exactly one of these transition metal atoms was substituted by each of the respective dopants Cr, Ni and Co, with the remaining 31 transition metal atoms being Fe. The calculated densities-of-states were performed on a relatively tight energy grid of 0.3 mRyd (4.08 meV) and a Gaussian smearing parameter of 0.5 mRyd was used. Convergence checks comparing calculated densities-of-states with 300 and 1000 k-points found scant differences so that the latter number is considered sufficient to describe the properties.

## Data Availability

The data sets that support the findings in this study are available from the corresponding author upon request.

## Code availability

The codes that support the findings in this study are available from the corresponding author upon request.

## Acknowledgments

This research was conducted at the Center for Nanophase Materials Sciences, which is a DOE Office of Science User Facility. L.L., A.S., D.S.P. and Z.G.'s research were supported by the U.S. DOE, Office of Science, Basic Energy Sciences, Materials Sciences and Engineering Division.

## Competing interests

384 The authors declare no competing interests.

## 385 Author contributions

386 Q.Z., M.F, Z.W. and Z.G. conducted the STM/S studies and data analysis; L.L. and A.S.
387 grew the crystals; D.P. did the DFT calculations; all authors contributed to the discussion and
388 writing of the manuscript.

## 389 Supporting information

## 390 Reference

391 1 Lee, P. A., Nagaosa, N. & Wen, X.-G. Doping a Mott insulator: Physics of high-temperature
392 superconductivity. *Reviews of Modern Physics* **78**, 17-85, doi:10.1103/RevModPhys.78.17 (2006).
393 2 Dai, P. Antiferromagnetic order and spin dynamics in iron-based superconductors. *Reviews of*
394 *Modern Physics* **87**, 855-896, doi:10.1103/RevModPhys.87.855 (2015).
395 3 Hashimoto, M., Vishik, I. M., He, R.-H., Devereaux, T. P. & Shen, Z.-X. Energy gaps in high-
396 transition-temperature cuprate superconductors. *Nature Physics* **10**, 483,
397 doi:10.1038/nphys3009 (2014).
398 4 Scalapino, D. J. A common thread: The pairing interaction for unconventional superconductors.
399 *Reviews of Modern Physics* **84**, 1383-1417, doi:10.1103/RevModPhys.84.1383 (2012).
400 5 Tsuei, C. C. & Kirtley, J. R. Pairing symmetry in cuprate superconductors. *Reviews of Modern*
401 *Physics* **72**, 969-1016, doi:10.1103/RevModPhys.72.969 (2000).
402 6 Wang, F. & Lee, D.-H. The Electron-Pairing Mechanism of Iron-Based Superconductors. *Science*
403 **332**, 200, doi:10.1126/science.1200182 (2011).
404 7 Chubukov, A. Pairing Mechanism in Fe-Based Superconductors. *Annual Review of Condensed*
405 *Matter Physics* **3**, 57-92, doi:10.1146/annurev-conmatphys-020911-125055 (2012).
406 8 Bang, Y. & Stewart, G. R. Anomalous Scaling Relations and Pairing Mechanism of the Fe-based
407 Superconductors. *Journal of Physics: Conference Series* **807**, 052003, doi:10.1088/1742-
408 6596/807/5/052003 (2017).
409 9 Huang, D. & Hoffman, J. E. Monolayer FeSe on SrTiO3. *Annual Review of Condensed Matter Physics*
410 **8**, 311-336, doi:10.1146/annurev-conmatphys-031016-025242 (2017).
411 10 Rotter, M., Tegel, M. & Johrendt, D. Superconductivity at 38 K in the iron arsenide (Ba1-
412 xKx)Fe2As2. *Phys Rev Lett* **101**, 107006, doi:ARTN 107006

413 10.1103/PhysRevLett.101.107006 (2008).
414 11 Sefat, A. S. *et al.* BaT2As2 single crystals (T = Fe, Co, Ni) and superconductivity upon Co-doping.
415 *Physica C* **469**, 350-354, doi:10.1016/j.physc.2009.03.025 (2009).
416 12 Lu, X. Y. *et al.* Short-range cluster spin glass near optimal superconductivity in BaFe2-xNixAs2.
417 *Phys Rev B* **90**, 024509, doi:ARTN 024509

418 10.1103/PhysRevB.90.024509 (2014).

419 13 Jiang, S. *et al.* Superconductivity up to 30 K in the vicinity of the quantum critical point in
420 BaFe2(As1-xPx)(2). *J Phys-Condens Mat* **21**, 382203, doi:Artn 382203

421 10.1088/0953-8984/21/38/382203 (2009).
422 14 Sefat, A. S. *et al.* Absence of superconductivity in hole-doped BaFe2-xCrxAs2 single crystals. *Phys*
423 *Rev B* **79**, 224524, doi:ARTN 224524

424 10.1103/PhysRevB.79.224524 (2009).
425 15 Pratt, D. K. *et al.* Incommensurate Spin-Density Wave Order in Electron-Doped BaFe2As2
426 Superconductors. *Phys Rev Lett* **106**, 257001, doi:ARTN 257001

427 10.1103/PhysRevLett.106.257001 (2011).
428 16 Wang, F., Zhai, H., Ran, Y., Vishwanath, A. & Lee, D. H. Functional Renormalization-Group Study
429 of the Pairing Symmetry and Pairing Mechanism of the FeAs-Based High-Temperature
430 Superconductor. *Phys Rev Lett* **102**, 047005, doi:ARTN 047005

431 10.1103/PhysRevLett.102.047005 (2009).
432 17 Si, Q. M. & Abrahams, E. Strong correlations and magnetic frustration in the high T(c) iron
433 pnictides. *Phys Rev Lett* **101**, 076401, doi:ARTN 076401

434 10.1103/PhysRevLett.101.076401 (2008).
435 18 Kou, S. P., Li, T. & Weng, Z. Y. Coexistence of itinerant electrons and local moments in iron-based
436 superconductors. *EPL* **88**, 17010, doi:Artn 17010

437 10.1209/0295-5075/88/17010 (2009).
438 19 You, Y. Z., Yang, F., Kou, S. P. & Weng, Z. Y. Magnetic and superconducting instabilities in a hybrid
439 model of itinerant/localized electrons for iron pnictides. *Phys Rev B* **84**, 054527, doi:ARTN 054527

440 10.1103/PhysRevB.84.054527 (2011).
441 20 Zhou, X. D. *et al.* Evolution from Unconventional Spin Density Wave to Superconductivity and a
442 Pseudogaplike Phase in NaFe1-xCoxAs. *Phys Rev Lett* **109**, 037002, doi:ARTN 037002

443 10.1103/PhysRevLett.109.037002 (2012).
444 21 Bernhard, C. *et al.* Muon spin rotation study of magnetism and superconductivity in Ba(Fe1-
445 xCox)(2)As-2 single crystals. *Phys Rev B* **86**, 184509, doi:ARTN 184509

446 10.1103/PhysRevB.86.184509 (2012).
447 22 Bernhard, C. *et al.* Muon spin rotation study of magnetism and superconductivity in BaFe2-
448 xCoxAs2 and Pr1-xSrxFeAsO. *New J Phys* **11**, 055050, doi:Artn 055050

449 10.1088/1367-2630/11/5/055050 (2009).
450 23 Cai, P. *et al.* Visualizing the microscopic coexistence of spin density wave and superconductivity
451 in underdoped NaFe1-xCoxAs. *Nat Commun* **4**, 1596, doi:ARTN 1596

452 10.1038/ncomms2592 (2013).
453 24 Sefat, A. S. Bulk synthesis of iron-based superconductors. *Curr Opin Solid St M* **17**, 59-64,
454 doi:10.1016/j.cossms.2013.04.001 (2013).
455 25 Zou, Q. *et al.* Effect of Surface Morphology and Magnetic Impurities on the Electronic Structure in
456 Cobalt-Doped BaFe2As2 Superconductors. *Nano Lett* **17**, 1642-1647 (2017).
457 26 Hoffman, J. E. Spectroscopic scanning tunneling microscopy insights into Fe-based
458 superconductors. *Rep Prog Phys* **74**, 124513, doi:Artn 124513

459 10.1088/0034-4885/74/12/124513 (2011).

460 27 Somnath, S., Smith, C. R., Jesse, S. & Laanait, N. Pycroscopy - An Open Source Approach to
461 Microscopy and Microanalysis in the Age of Big Data and Open Science. *Microscopy and*
462 *Microanalysis* **23**, 224-225, doi:10.1017/S1431927617001805 (2017).
463 28 Pan, S. H. *et al.* STM studies of the electronic structure of vortex cores in Bi2Sr2CaCu2O8+delta.
464 *Phys Rev Lett* **85**, 1536-1539, doi:DOI 10.1103/PhysRevLett.85.1536 (2000).
465 29 Hanaguri, T. *et al.* Scanning tunneling microscopy/spectroscopy of vortices in LiFeAs. *Phys Rev B*
466 **85**, 214505, doi:ARTN 214505

467 10.1103/PhysRevB.85.214505 (2012).
468 30 Yin, Y. *et al.* Scanning tunneling spectroscopy and vortex imaging in the iron pnictide
469 superconductor BaFe1.8Co0.2As2. *Phys Rev Lett* **102**, 097002,
470 doi:10.1103/PhysRevLett.102.097002 (2009).
471 31 Shan, L. *et al.* Observation of ordered vortices with Andreev bound states in Ba0.6K0.4Fe2As2.
472 *Nature Physics* **7**, 325-331, doi:10.1038/Nphys1908 (2011).
473 32 Yin, J. *et al.* Observation of a robust zero-energy bound state in iron-based superconductor Fe (Te,
474 Se). *Nature Physics* **11**, 543 (2015).
475 33 Li, W. *et al.* Phase separation and magnetic order in K-doped iron selenide superconductor. *Nature*
476 *Physics* **8**, 126 (2012).
477 34 Gastiasoro, M. N., Hirschfeld, P. J. & Andersen, B. M. Impurity states and cooperative magnetic
478 order in Fe-based superconductors. *Phys Rev B* **88**, 220509(R), doi:ARTN 220509

479 10.1103/PhysRevB.88.220509 (2013).
480 35 P. Blaha, K. S., G. K. H. Madsen, D. Kvasnicka, J. Luitz, R. Laskowski, F. Tran and L. D. Marks. *WIEN2k,*
481 *An Augmented Plane Wave + Local Orbitals Program for Calculating Crystal Properties*. (Techn.
482 Universität Wien, Austria, 2018).
483 36 Perdew, J. P., Burke, K. & Ernzerhof, M. Generalized gradient approximation made simple. *Phys*
484 *Rev Lett* **77**, 3865-3868, doi:DOI 10.1103/PhysRevLett.77.3865 (1996).
485 37 Avci, S. *et al.* Phase diagram of Ba1-xKxFe2As2. *Phys Rev B* **85**, doi:ARTN 184507

486 10.1103/PhysRevB.85.184507 (2012).

# Figure Legends

Fig. 1. Morphology and electronic structure of $Ba(Fe_{0.96}Ni_{0.04})_2As_2$. (a) The phase diagram of $BaFe_2As_2$ under
doping concentration variation. The inset is the crystal structure of the parent compound $BaFe_2As_2$. (b) The large
scale STM topography image of Ni-122, $V_{Bias}$ = -20 mV, $I_t$ = 100 pA. The red and green boxes mark the locations of
(c) and Fig. 2 (a), respectively. The scale bar represents 30 nm. (c) Atomically resolved STM image of Ni-122
surface ($V_{Bias}$ = -100 mV, $I_t$ = 400 pA) shows $\sqrt{2} \times \sqrt{2}$ reconstruction, the scale bar represents 3 nm. (d) Sets of STS
spectra along the black arrow in (c), show local electronic inhomogeneity. The blue and brown curves show
superconducting and in-gap states as labeled. Curves are shifted vertically for clarity.

Fig. 2 Electronic inhomogeneity on the surface of $Ba(Fe_{0.96}Ni_{0.04})_2As_2$. (a) The STM topography image of Ni-122 from
the green box in Fig. 1(a), $V_{Bias}$ = -20 mV and $I_t$ = 100 pA, the scale bar represents 6 nm. (b) STS map at Fermi level
(0 meV) and (c) superconducting gap map from the same area (blue box in (a)), the scale bar represents 3 nm. (d) and
(e) The line STS spectra along the brown and green arrows in (b). Brown and green circles in (b) and spectra in (d)
and (e) outline the non-SC areas. (f) Three K-means principal responses and (g) cluster map which shows the spatial
distribution of each type of spectra from the same area as (b).

Fig. 3 K-means clustering comparison of Ni-122, Co-122 and Cr-122. STM topographies of Ni-122 (a), Co-122 (d)
and Cr-122 (g). (b), (e) and (h) Cluster maps constructed from CITS simultaneously collected with (a), (d) and (g).
(c), (f) and (i) K-means principal responses of (b), (e) and (h). (j) The area percentages of the principal responses in
Ni-122, Co-122 and Cr-122 from the three images in (a), (d) and (g). (k) LDOS comparison of the in-gap states of Ni-
122 and Co-122. Ni-122, 130 nm x 130 nm, 256 x 256 pixels; Co-122, 104 nm x 104 nm, 128 x 128 pixels; Cr-122,
104 nm x 104 nm, 256 x 256 pixels, the scale bars represent 25 nm.

Fig. 4 Vortex imaging in Ni-122 (a), (b) and (c) STS maps at $E_F$ show the vortex lattices from the same location at
various vertical magnetic field. ($V_{Bias}$ = -10 mV, $I_t$ = 100 pA and $T$ = 4.2 K). (d), (e) and (f) Two-dimensional self-
correlation images calculated from vortex images shown in (a), (b) and (c). (g), (h), and (i) The FFT of the vortex
images. (j), (k) and (l) Delaunay triangulation diagrams overlaid on vortex images shown in (a), (b) and (c). Triangles
emphasize the vortices with nearest neighbors different from 6 (cyan: 3; yellow: 4; blue: 5 and red: 7). (m) Spatial
evolution of STS crossing a vortex core center along the black arrow in the inset. The spectra are offset for clarity. (n)
Magnetic field dependent vortex density in Ni-122, the error bars represent the standard deviation. (o) The fractions
of defect-vortex at different magnetic fields for Ni-122 and Co-122, the scale bars represent 40 nm for (a)-(f), and (j)-
(l), 0.05 $nm^{-1}$ for (g) to (i).

Fig. 5 Magnetic field dependence of the spectra of Co-122. (a), (b), (c) and (d) SC, in-gap, S-shape and L-shape
spectra measured in a magnetic field of 0 - 6 Tesla at 2.0 K. The dashed lines are guides for the eye. Each set of field
dependent spectra is collected at the same atomic locations; the curves in (b), (c) and (d) are offset vertically for
clarity.

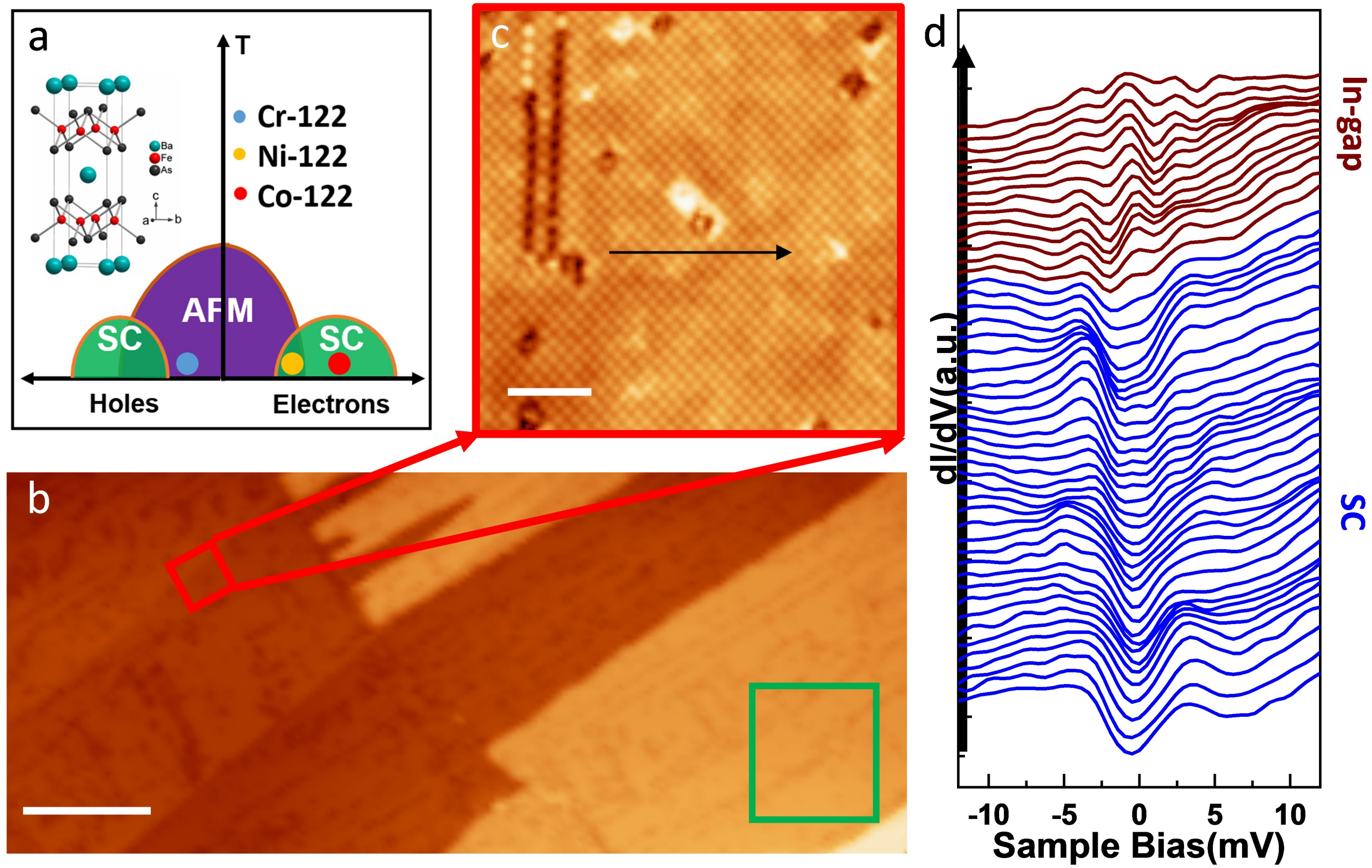

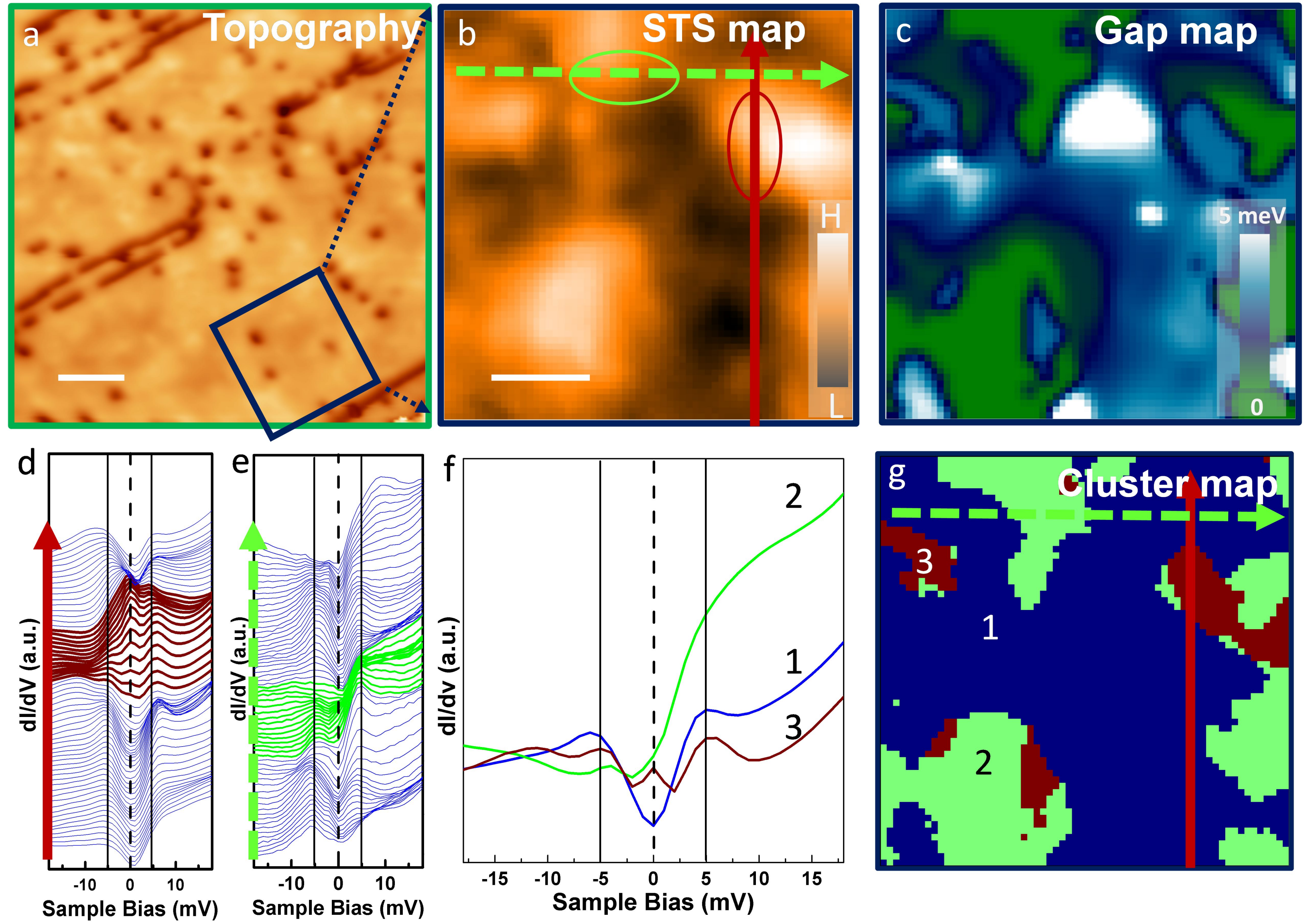

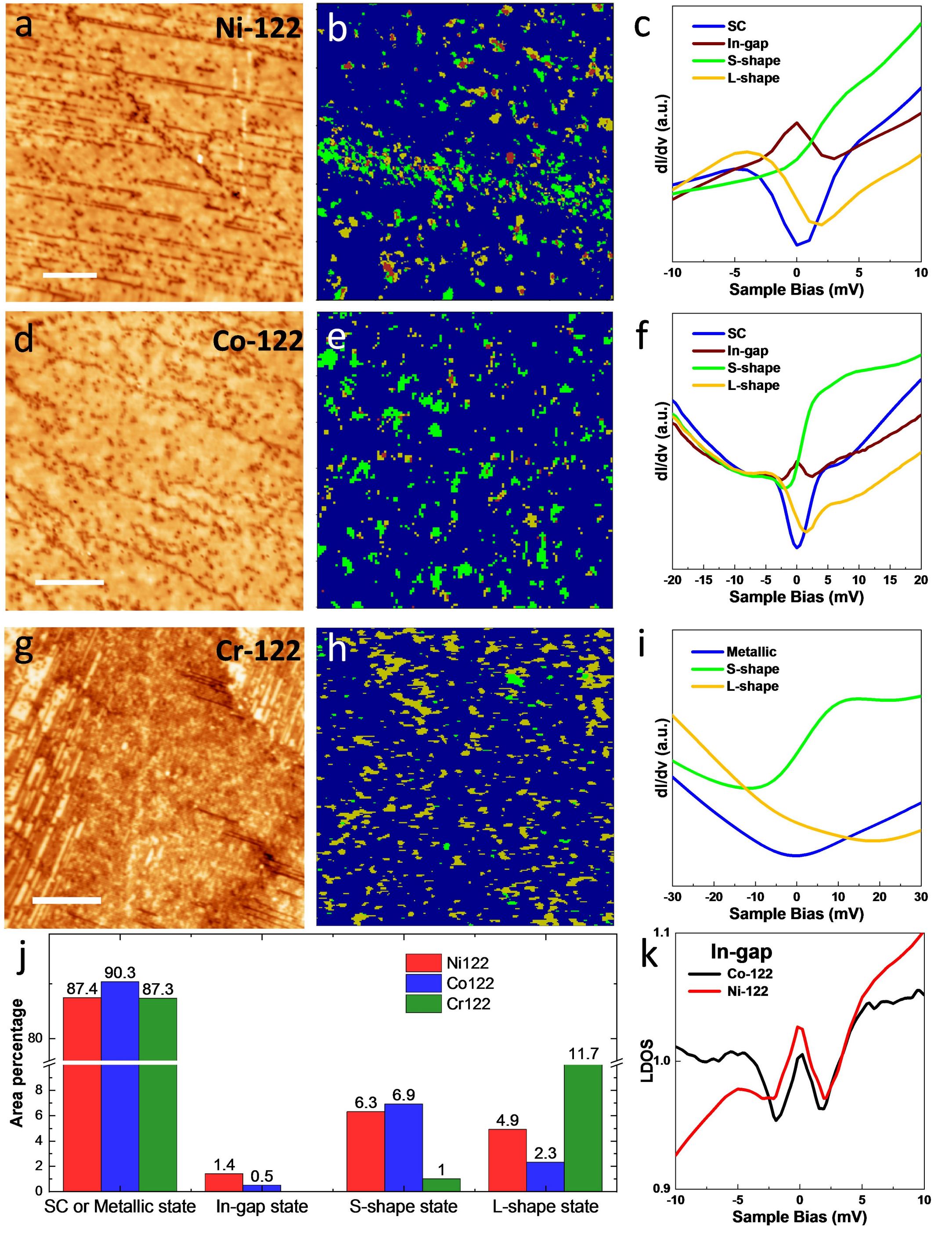

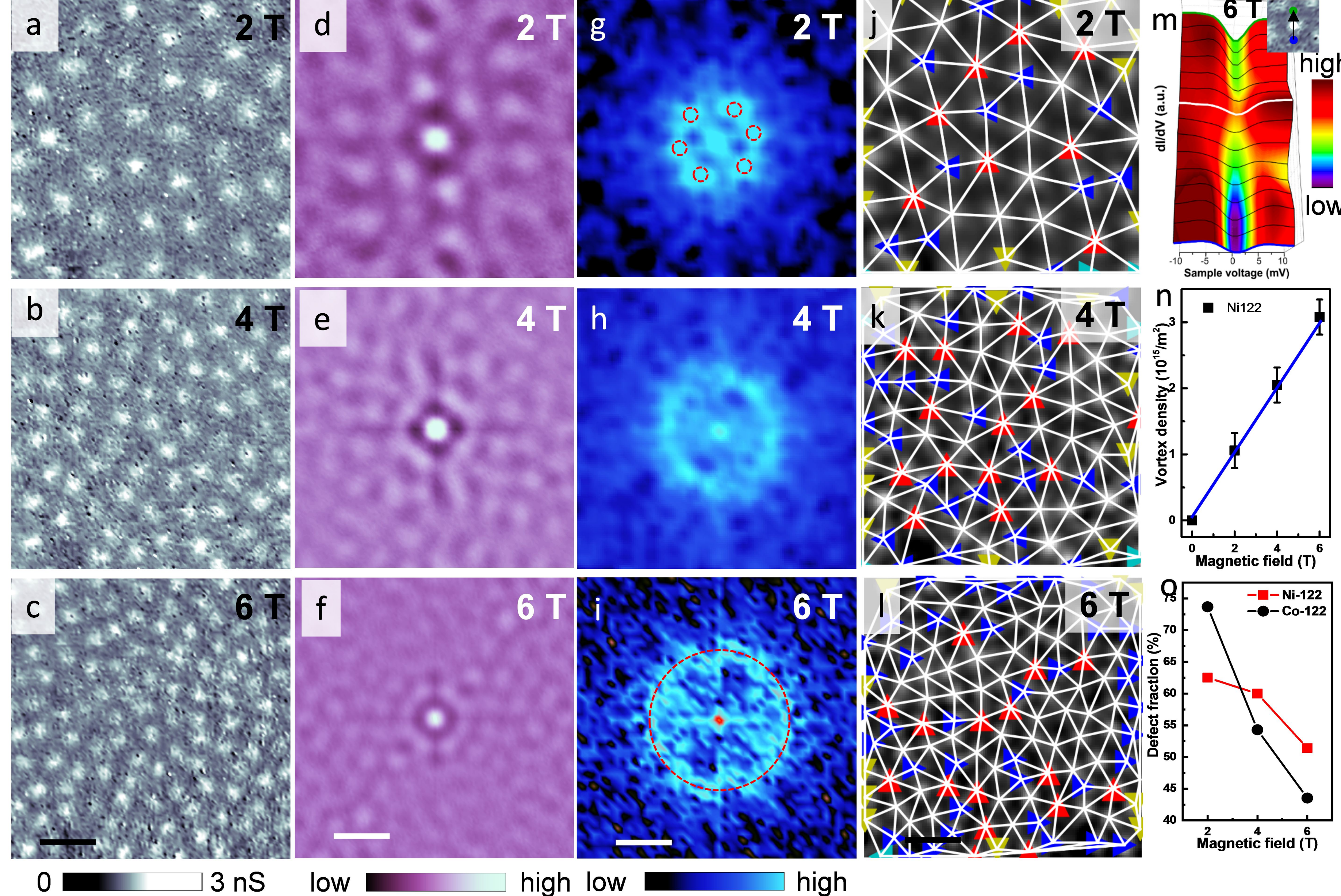

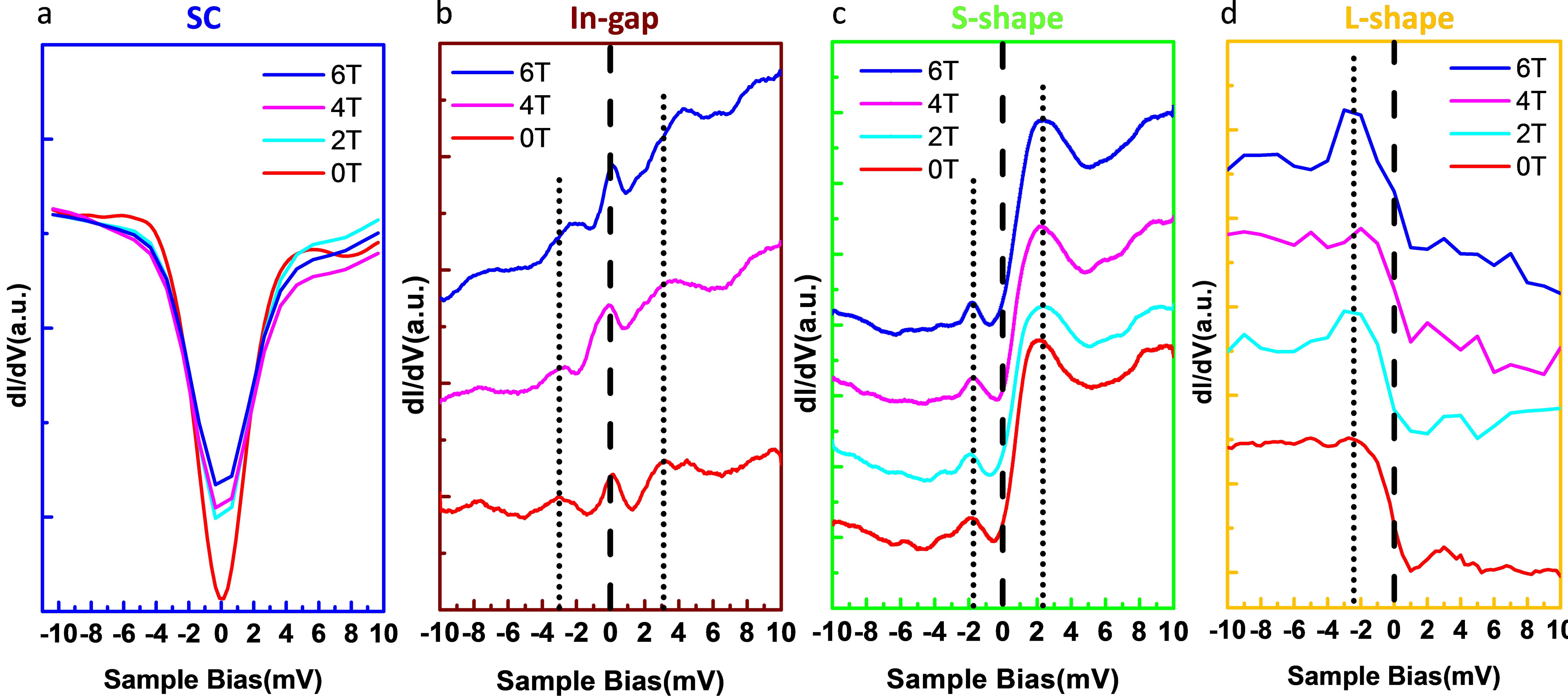